\documentclass[superscriptaddress,nofootinbib,a4paper,11pt]{article}
\usepackage{jheppub}
\usepackage{graphicx}
\usepackage{subcaption}
\usepackage{float}
\usepackage{amsmath}
\usepackage{amssymb}
\usepackage[utf8]{inputenc}
\usepackage{hyperref}
\usepackage{color}
\usepackage{slashed}
\usepackage{multirow}
\usepackage{enumitem}
\usepackage{lineno}
\allowdisplaybreaks

\begin{document}

\title{\boldmath Bypassing Spin-Analyzing Power Dependence for Quantum Entanglement at Colliders: A Case Study of $\Lambda\bar{\Lambda}$
}

\author[a]{Junle Pei,}
\author[b]{Tianjun Li,}
\author[c,*]{Lina Wu,\note[*]{Corresponding author.}}
\author[b]{Xiqing Hao,}
\author[b]{Xiaochuan Wang}

\affiliation[a]{Institute of Physics, Henan Academy of Sciences, Zhengzhou 450046, P. R. China}
\affiliation[b]{School of Physics, Henan Normal University, Xinxiang 453007, P. R. China}
\affiliation[c]{School of Sciences, Xi'an Technological University, Xi'an 710021, P. R. China}

\emailAdd{peijunle@hnas.ac.cn}
\emailAdd{tli@itp.ac.cn}
\emailAdd{wulina@xatu.edu.cn}
\emailAdd{haoxiqing@htu.edu.cn}
\emailAdd{xcwang@htu.edu.cn}

\abstract{
We study, as a concrete case study using the $\Lambda(\to p\pi^-)\bar{\Lambda}(\to \bar{p}\pi^+)$ system, whether quantum entanglement in fermion pairs produced at colliders can be certified solely using angular information from final-state decays, while remaining independent of the parity-violating decay parameters $\alpha_\Lambda$ and $\alpha_{\bar{\Lambda}}$. Building on a general decomposition of any angular observable in terms of Wigner d-functions, we show that the expectation value must take the form $\mathcal{O}_0+\mathcal{O}_1\alpha_\Lambda+\mathcal{O}_2\alpha_{\bar{\Lambda}}+\mathcal{O}_3\alpha_\Lambda\alpha_{\bar{\Lambda}}$, with coefficients $\mathcal{O}_i$ ($i=0,1,2,3$) linear in the spin-density matrix elements $\alpha_{k,j}\alpha^*_{m,n}$. We obtain the value ranges of observables over the general and separable spaces of $\alpha_{k,j}$, and demonstrate a sufficient entanglement condition for pure states, extending it to mixed states by convexity. In constructing an $\alpha_\Lambda$- and $\alpha_{\bar{\Lambda}}$-independent witness from angular observables alone, we find that there are obstacles to probe quantum entanglement via the inequality-type and ratio-type ways. In particular, for the ratio-type criterion ${\langle A\rangle}/{\langle B\rangle}$, the presence of zeros of $\langle B\rangle$ in both the general and separable spaces of $\alpha_{k,j}~(k,j=\pm\frac{1}{2})$ results in identical value ranges of ${\langle A\rangle}/{\langle B\rangle}$ in the two spaces (covering the entire real line), thereby precluding any effective criterion.
Finally, for this specific system, we present the successful constructions with additional spin information: for the process of $e^+e^-\to J/\Psi\to \Lambda\bar{\Lambda}$ at an $e^+ e^-$ collider, 
independent spin information provided by beam-axis selection enables the construction of normalized observables $f_i~(i=1,2)$ that are insensitive to $\alpha_\Lambda$ and $\alpha_{\bar{\Lambda}}$; if their measured values lie in $\left[-1,-\tfrac{1}{2}\right)\cup\left(\tfrac{1}{2},1\right]$, entanglement is certified, irrespective of purity or mixedness.
}

\maketitle

\section{Introduction} \label{intro} 

Testing quantum mechanics and observing quantum entanglement (QE) in high-energy physics represent rapidly evolving frontiers \cite{BESIII:2018cnd,Afik:2020onf,Fabbrichesi:2021npl,Han:2023fci,ATLAS2024,Cheng:2023qmz,Subba:2024mnl,CMS:2024pts,Cheng:2024rxi,Subba:2024aut,Cheng:2025cuv,Han:2025ewp,vonKuk:2025kbv,Zhang:2025mmm,Pei:2025non,Pei:2025yvr,BESIII:2025vsr,Lin:2025eci,Wu:2025dds,Pacheco:2025qam}, where the interplay between quantum correlations and relativistic dynamics offers new tools for probing the foundations of physics. However, some studies \cite{ABEL1992304,Li:2024luk,Bechtle:2025ugc,Abel:2025skj,Low:2025aqq} contend that it is not possible to test locality via Bell's inequalities at colliders. Especially, the critiques \cite{Bechtle:2025ugc,Abel:2025skj} argue that current collider experiments can neither test whether nature is described by quantum mechanics or a local hidden-variable theory (LHVT), nor determine whether the polarization state of a multi-particle system is entangled or separable. They adhere to the credo that, if the purpose of an experiment is to test quantum mechanics, then no conclusion derived from quantum mechanics or quantum field theory (QFT) may be used at any stage of the analysis. In their view, present measurements rely on angular correlations among the momentum directions of different final-state particles to tomographically reconstruct correlations in the initial-state polarizations; however, the relation connecting the two (for example, in a fermion–antifermion system) involves the initial-state fermion's parameter governing parity (P) violation. Any independent determination of this parameter inevitably relies on theoretical calculations based on quantum mechanics or QFT. Consequently, the stated credo is not satisfied.

The central motivation of this work is to clarify whether QE in collider-produced fermion pairs such as $\Lambda\bar{\Lambda}$ can be certified using only angular information from decays $\Lambda\to p\pi^-$ and $\bar{\Lambda}\to \bar{p}\pi^+$, without relying on external knowledge of the parity-violating decay parameters $\alpha_\Lambda$ and $\alpha_{\bar{\Lambda}}$. This question is of both foundational and practical significance. Foundationally, the recent critiques have argued that any purported test of quantum mechanics must avoid theoretical inputs derived from quantum mechanics or QFT, which typically enter through the spin–analyzing powers of weak decays. Practically, uncertainties or potential biases in $\alpha_\Lambda$ and $\alpha_{\bar{\Lambda}}$ can obscure entanglement claims if these parameters are required as inputs. We therefore ask: can one construct an entanglement criterion that is strictly independent of $\alpha_\Lambda$ and $\alpha_{\bar{\Lambda}}$, using angular observables alone? We answer this question by analyzing the obstacles to constructing entanglement criteria independent of $\alpha_\Lambda$ and $\alpha_{\bar{\Lambda}}$ and presenting successful constructions with additional spin information in the $e^+e^-\to J/\Psi\to \Lambda\bar{\Lambda}$ process.

The paper is organized as follows. In Sec.~\ref{sec:2}, we set up the formalism by expanding angular observables in Wigner d-functions, deriving the decomposition $\mathcal{O}_0+\mathcal{O}_1\alpha_\Lambda+\mathcal{O}_2\alpha_{\bar{\Lambda}}+\mathcal{O}_3\alpha_\Lambda\alpha_{\bar{\Lambda}}$, and defining value ranges over general and separable spaces to formulate sufficient entanglement criteria for pure and mixed states.
In Sec.~\ref{sec:3}, we discuss the obstacles in formulating $\alpha_\Lambda$- and $\alpha_{\bar{\Lambda}}$-independent criteria of both inequality and ratio types. 
We present successful constructions with additional spin information in the $e^+e^-\to J/\Psi\to \Lambda\bar{\Lambda}$ process in Sec.~\ref{sec:4}. 
Finally, we conclude our study in Sec.~\ref{con}.

\section{Theory} \label{sec:2} 

Following Refs.~\cite{Pei:2025non,Pei:2025yvr}, we take $\Lambda(\to p\pi^-)\bar{\Lambda}(\to \bar{p}\pi^+)$ as an illustrative example.
For a $\Lambda\bar{\Lambda}$ pair in a pure state, its most general polarization state can be written in the helicity basis as
\begin{align}
    \left| \Lambda \bar{\Lambda}\right\rangle=\sum_{k,j=\pm\frac{1}{2}} \alpha_{k,j} \left|k\right\rangle_\Lambda \left|j\right\rangle_{\bar{\Lambda}}~,
\end{align}
where $k$ and $j$ denote the helicity quantum numbers of $\Lambda$ and $\bar{\Lambda}$, respectively, defined along their respective momentum directions in the center-of-mass (c.m.) frame.

For definiteness and to fix conventions, in the decays $\Lambda \to p+\pi^-$ and $\bar{\Lambda} \to \bar{p}+\pi^+$ we parameterize the unit momentum directions of $p$ in the $\Lambda$ rest frame and $\bar{p}$ in the $\bar{\Lambda}$ rest frame using spherical coordinates: $\hat{e}_p=(\sin\theta_1\cos\phi_1,\sin\theta_1\sin\phi_1,\cos\theta_1)$ and $\hat{e}_{\bar{p}}=(\sin\theta_2\cos\phi_2,\sin\theta_2\sin\phi_2,\cos\theta_2)$. The polar angles $\theta_1$ and $\theta_2$ are defined with respect to the $\Lambda$ momentum direction $\hat{e}_\Lambda$ in the $\Lambda\bar{\Lambda}$ c.m. frame. The azimuthal angles $\phi_1$ and $\phi_2$ ($\phi_i\in [0,~2\pi]$ for $i=1,2$) are measured from an arbitrary reference axis orthogonal to $\hat{e}_\Lambda$ and increase according to the right-hand rule about $\hat{e}_\Lambda$.
Then, any observable constructed from the angular variables can be expressed as \cite{Pei:2025yvr}
\begin{align}
    \langle\mathcal{O}(\theta_1,\theta_2,\phi_1,\phi_2)\rangle=\sum_{k,j,m,n=\pm \frac{1}{2}}
    \mathcal{O}_{k,j;m,n}\alpha_{k,j}\alpha_{m,n}^*
\end{align}
with
\begin{align}
   \mathcal{O}_{k,j;m,n}=&\frac{1}{16\pi^2}\sum_{\lambda_p,\lambda_{\bar{p}}=\pm \frac{1}{2}}\left(1-2\lambda_p\alpha_\Lambda\right)\left(1-2\lambda_{\bar{p}}\alpha_{\bar{\Lambda}}\right)
   \int_{-1}^1 d\cos\theta_1\int_{-1}^1 d\cos\theta_2\int_0^{2\pi}d\phi_1\int_0^{2\pi}d\phi_2\nonumber\\
  & \mathcal{O}(\theta_1,\theta_2,\phi_1,\phi_2) e^{i(k-m)\phi_1} e^{i(n-j)\phi_2}
   d^{\frac{1}{2}}_{k,\lambda_p}(\theta_1)
   d^{\frac{1}{2}}_{m,\lambda_p}(\theta_1)
   d^{\frac{1}{2}}_{\lambda_{\bar{p}},j}(\pi-\theta_2)
   d^{\frac{1}{2}}_{\lambda_{\bar{p}},n}(\pi-\theta_2)~.\label{anpdf}
\end{align}
Here, \(\lambda_{p}\) and \(\lambda_{\bar{p}}\) ($\lambda_p,\lambda_{\bar{p}}=\pm\frac{1}{2}$) are spin projections of $p$ and $\bar{p}$ defined relative to directions of \(\hat{e}_p\) and \(\hat{e}_{\bar{p}}\), respectively.
The explicit forms of the Wigner d-functions involved are given by
\begin{align}
    d^{\frac{1}{2}}_{\frac{1}{2},\frac{1}{2}}(\theta)=d^{\frac{1}{2}}_{-\frac{1}{2},-\frac{1}{2}}(\theta)=\cos\frac{\theta}{2}~,\quad
    d^{\frac{1}{2}}_{-\frac{1}{2},\frac{1}{2}}(\theta)=-d^{\frac{1}{2}}_{\frac{1}{2},-\frac{1}{2}}(\theta)=\sin\frac{\theta}{2}~.
\end{align}
Consequently, $\langle\mathcal{O}(\theta_1,\theta_2,\phi_1,\phi_2)\rangle$ can also be written as
\begin{align}
    \langle\mathcal{O}(\theta_1,\theta_2,\phi_1,\phi_2)\rangle=\mathcal{O}_0+\mathcal{O}_1\alpha_\Lambda+\mathcal{O}_2\alpha_{\bar{\Lambda}}+\mathcal{O}_3\alpha_\Lambda\alpha_{\bar{\Lambda}}~,
\end{align}
where
\begin{align}
    \mathcal{O}_i=\sum_{k,j,m,n=\pm \frac{1}{2}} \mathcal{O}^i_{k,j;m,n}\alpha_{k,j}\alpha_{m,n}^*~,\quad i=0,1,2,3~.
\end{align}
A straightforward calculation yields
\begin{align}
    \mathcal{O}^0_{k,j;m,n}=&
    \frac{1}{16\pi^2}
   \int_{-1}^1 d\cos\theta_1\int_{-1}^1 d\cos\theta_2\int_0^{2\pi}d\phi_1\int_0^{2\pi}d\phi_2~ \mathcal{O}(\theta_1,\theta_2,\phi_1,\phi_2) \delta^k_m \delta^j_n~, \label{o0}\\
    \mathcal{O}^1_{k,j;m,n}=&-
    \frac{1}{16\pi^2}
   \int_{-1}^1 d\cos\theta_1\int_{-1}^1 d\cos\theta_2\int_0^{2\pi}d\phi_1\int_0^{2\pi}d\phi_2~ \mathcal{O}(\theta_1,\theta_2,\phi_1,\phi_2)\nonumber\\
   &\times e^{i(k-m)\phi_1}\sin\left(\theta_1+\frac{\pi}{2}(k+m)\right) \delta^j_n~, \label{o1}\\
    \mathcal{O}^2_{k,j;m,n}=&
    \frac{1}{16\pi^2}
   \int_{-1}^1 d\cos\theta_1\int_{-1}^1 d\cos\theta_2\int_0^{2\pi}d\phi_1\int_0^{2\pi}d\phi_2~ \mathcal{O}(\theta_1,\theta_2,\phi_1,\phi_2)\nonumber\\
   & \times e^{i(n-j)\phi_2}\sin\left(\theta_2+\frac{\pi}{2}(j+n)\right) \delta^k_m~, \label{o2}\\
   \mathcal{O}^3_{k,j;m,n}=&
    -\frac{1}{16\pi^2}
   \int_{-1}^1 d\cos\theta_1\int_{-1}^1 d\cos\theta_2\int_0^{2\pi}d\phi_1\int_0^{2\pi}d\phi_2~ \mathcal{O}(\theta_1,\theta_2,\phi_1,\phi_2)\nonumber\\
   & \times e^{i(k-m)\phi_1}e^{i(n-j)\phi_2}\sin\left(\theta_1+\frac{\pi}{2}(k+m)\right)\sin\left(\theta_2+\frac{\pi}{2}(j+n)\right)~. \label{o3}
\end{align}

The general space of $\alpha_{k,j}~(k,j=\pm\frac{1}{2})$ is defined by
\begin{align}
     \sum_{k,j=\pm\frac{1}{2}} \left|\alpha_{k,j}\right|^2=1~,
\end{align}
whereas the separable space of $\alpha_{k,j}$ where there is no QE is defined by
\begin{align}
    \alpha_{k,j}=\beta_k \gamma_j~,\quad \sum_{k=\pm\frac{1}{2}}\left|\beta_k\right|^2=\sum_{j=\pm\frac{1}{2}}\left|\gamma_j\right|^2=1~.
\end{align}
Let the value ranges of $\langle\mathcal{O}(\theta_1,\theta_2,\phi_1,\phi_2)\rangle$ over the general space and the separable space of $\alpha_{k,j}~(k,j=\pm\frac{1}{2})$ be denoted by $R_1$ and $R_2$, respectively. Since both the general and separable spaces of $\alpha_{k,j}$ are connected, and each $\mathcal{O}_i$ is a linear combination of $\alpha_{k,j}\alpha_{m,n}^*~(k,j,m,n=\pm\frac{1}{2})$, it follows that $R_1$ and $R_2$ are themselves connected. Therefore, if the measurement yields $\langle\mathcal{O}(\theta_1,\theta_2,\phi_1,\phi_2)\rangle\in R_1\backslash R_2$, this serves as a sufficient condition for the presence of QE in the $\Lambda\bar{\Lambda}$ system.
It can be shown that the entanglement criterion derived above for pure states applies equally to mixed states:
\begin{align}
    \langle\mathcal{O}\rangle\in R_k \Longrightarrow \sum_{i} p_i\langle\mathcal{O}\rangle_i\in R_k~,\quad k=1,2~,
\end{align}
where $p_i$ denotes the probability weight of the $i$-th component in the mixture, satisfying $0< p_i<1$ and $\sum_i p_i=1$.

\section{Obstacles to Constructing $\alpha_\Lambda$- and $\alpha_{\bar{\Lambda}}$-Independent Criteria} \label{sec:3} 

Our analysis is based on tomographically reconstructing the spin–polarization information of the $\Lambda\bar{\Lambda}$ system from measurements of angular observables. Therefore, the most primitive and complete information available is the angular distribution of the final-state particles, denoted by $\mathcal{W}(\theta_1,\theta_2,\phi_1,\phi_2)$.
According to Eq.~(\ref{anpdf}), the explicit form of $\mathcal{W}(\theta_1,\theta_2,\phi_1,\phi_2)$ is given by
\begin{align}
   \mathcal{W}(\theta_1,\theta_2,\phi_1,\phi_2)=&\frac{1}{16\pi^2}\sum_{\lambda_p,\lambda_{\bar{p}}=\pm \frac{1}{2}}\left(1-2\lambda_p\alpha_\Lambda\right)\left(1-2\lambda_{\bar{p}}\alpha_{\bar{\Lambda}}\right) \sum_{k,j,m,n=\pm \frac{1}{2}} \alpha_{k,j}\alpha_{m,n}^*\times
    \nonumber\\
  & e^{i(k-m)\phi_1} e^{i(n-j)\phi_2}
   d^{\frac{1}{2}}_{k,\lambda_p}(\theta_1)
   d^{\frac{1}{2}}_{m,\lambda_p}(\theta_1)
   d^{\frac{1}{2}}_{\lambda_{\bar{p}},j}(\pi-\theta_2)
   d^{\frac{1}{2}}_{\lambda_{\bar{p}},n}(\pi-\theta_2)~.
\end{align}
An entanglement criterion independent of $\alpha_\Lambda$ and $\alpha_{\bar{\Lambda}}$ means that varying the values of $\alpha_\Lambda$ and $\alpha_{\bar{\Lambda}}$ does not change the verdict. By setting $\alpha_\Lambda$ and $\alpha_{\bar{\Lambda}}$ to zero, one can obtain a part of $\mathcal{W}(\theta_1,\theta_2,\phi_1,\phi_2)$ that does not change with $\alpha_\Lambda$ and $\alpha_{\bar{\Lambda}}$ (denoted by $\mathcal{W^\prime}(\theta_1,\theta_2,\phi_1,\phi_2)$), which is
\begin{align}
   \mathcal{W^\prime}(\theta_1,\theta_2,\phi_1,\phi_2)=&\frac{1}{16\pi^2}~.
\end{align}
Since $\mathcal{W^\prime}(\theta_1,\theta_2,\phi_1,\phi_2)$ contains no information about $\alpha_{k,j}~(k,j=\pm\frac{1}{2})$, it is impossible to extract any information about the initial system's spin polarization from angular observables when there is no violation of parity ($\alpha_\Lambda=\alpha_{\bar{\Lambda}}=0$). 
Accordingly, for $\langle\mathcal{O}(\theta_1,\theta_2,\phi_1,\phi_2)\rangle$ that does not depend on $\alpha_\Lambda$ and $\alpha_{\bar{\Lambda}}$, from Eq.~(\ref{o0}) it follows that
\begin{align}
    \langle\mathcal{O}(\theta_1,\theta_2,\phi_1,\phi_2)\rangle=\mathcal{O}_0=\frac{1}{16\pi^2}
   \int_{-1}^1 d\cos\theta_1\int_{-1}^1 d\cos\theta_2\int_0^{2\pi}d\phi_1\int_0^{2\pi}d\phi_2~ \mathcal{O}(\theta_1,\theta_2,\phi_1,\phi_2)~.
\end{align}
Since $\mathcal{O}_0$ does not depend on $\alpha_{k,j}~(k,j=\pm\frac{1}{2})$, $\langle\mathcal{O}(\theta_1,\theta_2,\phi_1,\phi_2)\rangle$ cannot provide a QE criterion. 

We are therefore led to consider a more general construction:
\begin{align}
f\left(\langle\mathcal{D}_1\rangle,\langle\mathcal{D}_2\rangle,...,\langle \mathcal{D}_N\rangle\right)~,
\end{align}
where $f$ is a function of the expectation values of multiple angular observables $\langle\mathcal{D}_i\rangle~(i=1,2,...,N)$. To obtain an entanglement criterion that is independent of $\alpha_\Lambda$ and $\alpha_{\bar{\Lambda}}$, $f$ must satisfy the following conditions:
\begin{itemize}
    \item $f$ itself is independent of $\alpha_\Lambda$ and $\alpha_{\bar{\Lambda}}$. 
    \item For both pure states and arbitrary mixed states, the set difference between the value ranges of $f$ over the general and separable spaces of $\alpha_{k,j}~(k,j=\pm\frac{1}{2})$ is nonempty.
\end{itemize}

\subsection{Inequality-Type Criteria} \label{sec:3-1} 

If $\mathcal{O}_p$ and $\mathcal{O}_{\bar{p}}$ depend solely on the angular variables of $p$ ($\theta_1$ and $\phi_1$) and $\bar{p}$ ($\theta_2$ and $\phi_2$), respectively, then for pure states there exists a sufficient condition for QE of the form
\begin{align}
\langle\mathcal{O}_p\mathcal{O}_{\bar{p}}\rangle\ne\langle\mathcal{O}_p\rangle\langle\mathcal{O}_{\bar{p}}\rangle~.\label{llb}
\end{align} 
For example, 
\begin{align}
    &\langle\cos\theta_1\rangle=\frac{\alpha_\Lambda}{3}
    \left(\left|\alpha_{-\frac{1}{2},-\frac{1}{2}}\right|^2+\left|\alpha_{-\frac{1}{2},\frac{1}{2}}\right|^2-\left|\alpha_{\frac{1}{2},-\frac{1}{2}}\right|^2-\left|\alpha_{\frac{1}{2},\frac{1}{2}}\right|^2\right)~, \\
    &\langle\cos\theta_2\rangle=-\frac{\alpha_{\bar{\Lambda}}}{3}
    \left(\left|\alpha_{-\frac{1}{2},-\frac{1}{2}}\right|^2-\left|\alpha_{-\frac{1}{2},\frac{1}{2}}\right|^2+\left|\alpha_{\frac{1}{2},-\frac{1}{2}}\right|^2-\left|\alpha_{\frac{1}{2},\frac{1}{2}}\right|^2\right)~,\\
    &\langle\cos\theta_1\cos\theta_2\rangle=-\frac{\alpha_\Lambda \alpha_{\bar{\Lambda}}}{9}
    \left(\left|\alpha_{-\frac{1}{2},-\frac{1}{2}}\right|^2-\left|\alpha_{-\frac{1}{2},\frac{1}{2}}\right|^2-\left|\alpha_{\frac{1}{2},-\frac{1}{2}}\right|^2+\left|\alpha_{\frac{1}{2},\frac{1}{2}}\right|^2\right)~.
\end{align}
The inequality-type criterion in Eq.~(\ref{llb}) can be made independent of $\alpha_\Lambda$ and $\alpha_{\bar{\Lambda}}$.
Observing a pure state that satisfies Eq.~(\ref{llb}) allows one to conclude the presence of QE without knowing the specific values of $\alpha_\Lambda$ and $\alpha_{\bar{\Lambda}}$.

However, for a two-component mixed state in which each component is unentangled, one finds
\begin{align}
& \langle\mathcal{O}_p\rangle\langle\mathcal{O}_{\bar{p}}\rangle=\left(p_1\langle\mathcal{O}_p\rangle_1+p_2\langle\mathcal{O}_p\rangle_2\right)\left(p_1\langle\mathcal{O}_{\bar{p}}\rangle_1+p_2\langle\mathcal{O}_{\bar{p}}\rangle_2\right)~,\\
& \langle\mathcal{O}_p\mathcal{O}_{\bar{p}}\rangle=
p_1 \langle\mathcal{O}_p\mathcal{O}_{\bar{p}}\rangle_1+p_2 \langle\mathcal{O}_p\mathcal{O}_{\bar{p}}\rangle_2=
p_1 \langle\mathcal{O}_p\rangle_1\langle\mathcal{O}_{\bar{p}}\rangle_1+p_2 \langle\mathcal{O}_p\rangle_2\langle\mathcal{O}_{\bar{p}}\rangle_2~.
\end{align}
In this case, $\langle\mathcal{O}_p\rangle\langle\mathcal{O}_{\bar{p}}\rangle$ and $\langle\mathcal{O}_p\mathcal{O}_{\bar{p}}\rangle$ need not be equal. So, a mixed state satisfying Eq.~(\ref{llb}) is likewise unentangled (separable).

\subsection{Ratio-Type Criteria} \label{sec:3-2} 

For two distinct angular observables $A$ and $B$,
\begin{align}
    \frac{\langle A\rangle}{\langle B\rangle}=\frac{A_0+A_1\alpha_\Lambda+A_2\alpha_{\bar{\Lambda}}+A_3\alpha_{\Lambda}\alpha_{\bar{\Lambda}}}{B_0+B_1\alpha_\Lambda+B_2\alpha_{\bar{\Lambda}}+B_3\alpha_{\Lambda}\alpha_{\bar{\Lambda}}}~.
\end{align}
If ${\langle A\rangle}/{\langle B\rangle}$ is independent of $\alpha_\Lambda$ and $\alpha_{\bar{\Lambda}}$, there are two possibilities:
\begin{itemize}
    \item $A_0$ and $B_0$ are both nonzero. Since $A_0$ and $B_0$ do not depend on $\alpha_{k,j}~(k,j=\pm\frac{1}{2})$, in this case ${\langle A\rangle}/{\langle B\rangle}=A_0/B_0$ likewise does not depend on $\alpha_{k,j}$ and therefore cannot provide an entanglement criterion.
    \item $A_0$ and $B_0$ both vanish.
\end{itemize}

In the case that $A_0=B_0=0$, we analyze the zeros of $\langle B\rangle$ in the parameter space of $\alpha_{k,j}~(k,j=\pm\frac{1}{2})$. We select four points in the $\alpha_{k,j}$ space as follows:
\begin{align}
    P_1: \alpha_{-\frac{1}{2},-\frac{1}{2}}=1~,\quad
    P_2: \alpha_{-\frac{1}{2},\frac{1}{2}}=1~,\quad
    P_3: \alpha_{\frac{1}{2},-\frac{1}{2}}=1~,\quad
    P_4: \alpha_{\frac{1}{2},\frac{1}{2}}=1~.
\end{align}
Note that these four points exist in both the general and separable spaces of $\alpha_{k,j}$. At $P_1$, we write $\langle B\rangle$ as
\begin{align}
    \langle B\rangle_{P_1}=\tilde{B}_1\alpha_\Lambda+\tilde{B}_2\alpha_{\bar{\Lambda}}+\tilde{B}_3 \alpha_\Lambda \alpha_{\bar{\Lambda}}~,
\end{align}
where
\begin{align}
    \tilde{B}_1=&
    \frac{1}{16\pi^2}
   \int_{-1}^1 d\cos\theta_1\int_{-1}^1 d\cos\theta_2\int_0^{2\pi}d\phi_1\int_0^{2\pi}d\phi_2~ \mathcal{O}(\theta_1,\theta_2,\phi_1,\phi_2) \cos\theta_1~, \\
   \tilde{B}_2=&-
    \frac{1}{16\pi^2}
   \int_{-1}^1 d\cos\theta_1\int_{-1}^1 d\cos\theta_2\int_0^{2\pi}d\phi_1\int_0^{2\pi}d\phi_2~ \mathcal{O}(\theta_1,\theta_2,\phi_1,\phi_2) \cos\theta_2~, \\
   \tilde{B}_3=&-
    \frac{1}{16\pi^2}
   \int_{-1}^1 d\cos\theta_1\int_{-1}^1 d\cos\theta_2\int_0^{2\pi}d\phi_1\int_0^{2\pi}d\phi_2~ \mathcal{O}(\theta_1,\theta_2,\phi_1,\phi_2) \cos\theta_1 \cos\theta_2~.
\end{align}
Then, at the other points we have
\begin{align}
   & \langle B\rangle_{P_2}=\tilde{B}_1-\tilde{B}_2\alpha_{\bar{\Lambda}}-\tilde{B}_3 \alpha_\Lambda \alpha_{\bar{\Lambda}}~,\\
   & \langle B\rangle_{P_3}=-\tilde{B}_1+\tilde{B}_2\alpha_{\bar{\Lambda}}-\tilde{B}_3 \alpha_\Lambda \alpha_{\bar{\Lambda}}~,\\
   & \langle B\rangle_{P_4}=-\tilde{B}_1-\tilde{B}_2\alpha_{\bar{\Lambda}}+\tilde{B}_3 \alpha_\Lambda \alpha_{\bar{\Lambda}}~.
\end{align}
It follows that
\begin{align}
    \sum_{i=1}^4 \langle B\rangle_{P_i}=0~.
\end{align}
Since the general and separable spaces of $\alpha_{k,j}$ are each connected, and $B_i~(i=1,2,3)$ are linear combinations of $\alpha_{k,j}\alpha_{m,n}^*~(k,j,m,n=\pm \frac{1}{2})$, the above relation implies that zeros of $\langle B\rangle$ exist in both the general and separable spaces of $\alpha_{k,j}$. 
However, as we shall see next, the existence of zeros of $\langle B\rangle$ in the $\alpha_{k,j}$ parameter space poses obstacles to formulating QE criteria for both pure and mixed states.

We illustrate the obstruction for pure states with the following concrete example:
\begin{align}
    A=\cos(\phi_1+\phi_2)~,\quad \langle A\rangle=\kappa\left(\alpha_{-\frac{1}{2},\frac{1}{2}} \alpha^*_{\frac{1}{2},-\frac{1}{2}}+\alpha^*_{-\frac{1}{2},\frac{1}{2}} \alpha_{\frac{1}{2},-\frac{1}{2}} \right)~,\\
    B=\cos(\phi_1-\phi_2)~,\quad \langle B\rangle=\kappa\left(\alpha_{-\frac{1}{2},-\frac{1}{2}} \alpha^*_{\frac{1}{2},\frac{1}{2}}+\alpha^*_{-\frac{1}{2},-\frac{1}{2}} \alpha_{\frac{1}{2},\frac{1}{2}} \right)~,
\end{align}
where $\kappa=-\frac{\pi^2}{32}\alpha_\Lambda \alpha_{\bar{\Lambda}}$.
Value ranges of $\langle A\rangle/\kappa$, $\langle B\rangle/\kappa$, and $\langle A\rangle/\langle B\rangle$ for the pure state over the general and separable spaces of $\alpha_{k,j}~(k,j=\pm\frac{1}{2})$ are shown in Table \ref{emp1}, respectively.
\begin{table}[htb]
\centering
\begin{tabular}{cccc}
\hline
Observables & $R_1$ & $R_2$ & $R_1\backslash R_2$\\
\hline
$\langle A\rangle/\kappa$ &  $[-1,1]$ & $\left[-1/2,1/2\right]$ & $\left[-1,-1/2\right)\cup\left(1/2,1\right]$  \\
$\langle B\rangle/\kappa$ &  $[-1,1]$ & $\left[-1/2,1/2\right]$ & $\left[-1,-1/2\right)\cup\left(1/2,1\right]$ \\
$\langle A\rangle/\langle B\rangle$ &  $(-\infty,+\infty)$ & $(-\infty,+\infty)$ & $\emptyset$ \\
\hline
\end{tabular}
\caption{Value ranges of $\langle A\rangle/\kappa$, $\langle B\rangle/\kappa$, and $\langle A\rangle/\langle B\rangle$ for the pure state over the general and separable spaces of $\alpha_{k,j}~(k,j=\pm\frac{1}{2})$, respectively.}
\label{emp1}
\end{table}
In this example, because $\langle B\rangle$ has zeros, the value ranges of $\langle A\rangle/\langle B\rangle$ over the general and separable spaces of $\alpha_{k,j}$ (also denoted $R_1$ and $R_2$) both span the entire real line. Therefore, $\langle A\rangle/\langle B\rangle$ cannot be used to furnish a QE criterion.

To illustrate how the presence of zeros of $\langle B\rangle$ in the $\alpha_{k,j}$ parameter space obstructs establishing a QE criterion for mixed states, we consider the following hypothetical example:
\begin{align}
    \langle A^\prime\rangle=\kappa^\prime x^2~,\quad \langle B^\prime\rangle=\kappa^\prime x~.
\end{align}
Here, $\kappa^\prime$ depends on $\alpha_\Lambda$ and $\alpha_{\bar{\Lambda}}$, while $x$ depends on $\alpha_{k,j}$. For simplicity, we assume that the value ranges of $x$ over the general and separable spaces of $\alpha_{k,j}$ are $[-1,1]$ and $[-\frac{1}{2},\frac{1}{2}]$, respectively.
Thus, for pure states,
\begin{align}
\frac{\langle A^\prime\rangle}{\langle B^\prime\rangle}=x~.
\end{align}
This implies that the value ranges of $\langle A^\prime\rangle/\langle B^\prime\rangle$ over the general and separable spaces of $\alpha_{k,j}$ are $[-1,1]$ and $[-\frac{1}{2},\frac{1}{2}]$, respectively. Consequently, $\langle A^\prime\rangle/\langle B^\prime\rangle\in \left[-1,-\frac{1}{2}\right)\cup\left(\frac{1}{2},1\right]$ can serve as an entanglement criterion for pure states.
For a two-component mixed state, with the two components corresponding to $x_1$ and $x_2$, respectively, without loss of generality we take $x_1<x_2$. We then have
\begin{align}
\frac{\langle A^\prime\rangle}{\langle B^\prime\rangle}=\frac{ p_1 x_1^2+(1-p_1) x_2^2}{p_1 x_1+(1-p_1) x_2}~.
\end{align}
Letting $p_1$ vary over $(0,1)$ yields
\begin{align}
\frac{\langle A^\prime\rangle}{\langle B^\prime\rangle}=\begin{cases}
    (x_1,x_2)~,\quad x_1>0~\text{or}~ x_2<0  \\
    (-\infty,x_1)\cup(x_2,+\infty)~, \quad x_1<0<x_2 \\
    x_2~,\quad x_1=0 \\
    x_1~,\quad x_2=0
\end{cases}~.
\end{align}
By varying the values of $x_i~(i=1,2)$ and including the pure-state result, ${\langle A^\prime\rangle}/{\langle B^\prime\rangle}$ is found to span the entire real line over both the general and separable spaces of $\alpha_{k,j}$. Therefore, measurements of ${\langle A^\prime\rangle}/{\langle B^\prime\rangle}$ cannot be used to determine whether the system exhibits QE.

\section{Ratios with Additional Spin Information} \label{sec:4} 

From the discussion in Sec.~\ref{sec:3}, it is clear that there are obstacles to constructing entanglement criteria independent of $\alpha_\Lambda$ and $\alpha_{\bar{\Lambda}}$. In particular, for ratio-type criteria ${\langle A\rangle}/{\langle B\rangle}$, the presence of zeros of $\langle B\rangle$ in both the general and separable spaces of $\alpha_{k,j}~(k,j=\pm\frac{1}{2})$ forces ${\langle A\rangle}/{\langle B\rangle}$ to have identical value ranges in the two spaces (covering the entire real line), thereby precluding any effective criterion. However, by leveraging experimentally established information and restricting attention to regions of parameter space where $\langle B\rangle$ is zero-free, it remains possible to formulate viable entanglement criteria that are independent of $\alpha_\Lambda$ and $\alpha_{\bar{\Lambda}}$.

For on-shell $J/\Psi$ mesons produced at $e^+e^-$ colliders, experiments indicate that their spin projection quantum number along the beam axis is $\pm 1$. For the decay $J/\Psi\to \Lambda+\bar{\Lambda}$, if we restrict to $\Lambda\bar{\Lambda}$ pairs emitted back-to-back along the beam direction, then by angular-momentum conservation each event places the pair entirely in either $\left|-\frac{1}{2}\right\rangle_\Lambda\left|\frac{1}{2}\right\rangle_{\bar{\Lambda}}$ or $\left|\frac{1}{2}\right\rangle_\Lambda\left|-\frac{1}{2}\right\rangle_{\bar{\Lambda}}$. Consequently, for these events we obtain
\begin{align}
\langle\cos\theta_1\cos\theta_2\rangle_{\text{beam}}=\frac{\alpha_\Lambda \alpha_{\bar{\Lambda}}}{9}~.
\end{align}

Using the information provided by $\langle\cos\theta_1\cos\theta_2\rangle_{\text{beam}}$, we define
\begin{align}
    f_1=-\frac{32}{9\pi^2}\frac{\langle\cos(\phi_1+\phi_2)\rangle}{\langle\cos\theta_1\cos\theta_2\rangle_{\text{beam}}}=\left(\alpha_{-\frac{1}{2},\frac{1}{2}} \alpha^*_{\frac{1}{2},-\frac{1}{2}}+\alpha^*_{-\frac{1}{2},\frac{1}{2}} \alpha_{\frac{1}{2},-\frac{1}{2}} \right)~,\\
    f_2 =-\frac{32}{9\pi^2}\frac{\langle\cos(\phi_1+\phi_2)\rangle}{\langle\cos\theta_1\cos\theta_2\rangle_{\text{beam}}}=\left(\alpha_{-\frac{1}{2},-\frac{1}{2}} \alpha^*_{\frac{1}{2},\frac{1}{2}}+\alpha^*_{-\frac{1}{2},-\frac{1}{2}} \alpha_{\frac{1}{2},\frac{1}{2}} \right)~.
\end{align}
Therefore, for the process $e^+ + e^-\to J/\Psi\to \Lambda(\to p\pi^-)\Lambda(\to \bar{p}\pi^+)$, observing a value of $f_i~(i=1,2)$ within $\left[-1,-1/2\right)\cup\left(1/2,1\right]$ is sufficient to demonstrate the presence of QE in the $\Lambda\bar{\Lambda}$ pairs. Moreover, this conclusion is not only independent of whether the $\Lambda\bar{\Lambda}$ state is pure or mixed, but also does not depend on the values of $\alpha_\Lambda$ and $\alpha_{\bar{\Lambda}}$.

The successful construction above highlights a general principle recently articulated in Ref.~\cite{Cheng:2024rxi}: for quantum tomography at colliders, the required spin information can be extracted either from the decay (through spin-analyzing powers) or from the production kinematics. Here, we have effectively exchanged the dependence on the unknown decay parameters $\alpha_{\Lambda},\alpha_{\bar{\Lambda}}$ for a well-controlled production assumption--namely, that the $J/\psi$ mesons from $e^+e^-$ annihilation have spin projection quantum numbers $\pm 1$ along the beam axis, which is firmly established by experiment and angular momentum conservation. This is precisely the spirit of the “kinematic approach” of Ref.~\cite{Cheng:2024rxi} applied to the $\Lambda\bar{\Lambda}$ system. The two approaches (decay-based vs. production-based) are complementary; our ratio witnesses $f_1, f_2$ provide an example where production information allows an entanglement criterion that is completely independent of the parity-violating decay parameters, thereby circumventing the obstacles discussed in Sec.~\ref{sec:3}. We note that Ref.~\cite{Cheng:2024rxi} also emphasizes the superior statistical precision of the kinematic approach, a feature that may be worth exploring for hyperon pairs in future high-luminosity experiments.

\section{Conclusion} \label{con} 

We have systematically assessed the feasibility of certifying QE using only angular distributions in collider-produced two-fermion systems. We developed a unified formalism in which the expectation value of any angular observable is expressed as $\mathcal{O}_0+\mathcal{O}_1\alpha_\Lambda+\mathcal{O}_2\alpha_{\bar{\Lambda}}+\mathcal{O}_3\alpha_\Lambda\alpha_{\bar{\Lambda}}$, defined value ranges over the general and separable spaces of $\alpha_{k,j}~(k,j=\pm\frac{1}{2})$, and constructed sufficient criteria for pure states, extended to mixed states by convexity. We proved obstacles in formulating $\alpha_\Lambda$- and $\alpha_{\bar{\Lambda}}$-independent criteria of both inequality and ratio types: 
(i) For inequality-type criteria, obstructions arise when attempting to generalize from pure states to mixed states.  
(ii) For ratio-type criteria, the presence of zeros in the denominator can obstruct their applicability for both pure and mixed states.

At the same time, we outlined a constructive path with additional spin information: when experiments supply independent spin constraints tied to the beam axis (e.g., $e^+e^-\to J/\Psi\to \Lambda\bar{\Lambda}$), one can build normalized observables that are insensitive to $\alpha_\Lambda$ and $\alpha_{\bar{\Lambda}}$ and take distinct ranges on general versus separable spaces of $\alpha_{k,j}$, enabling direct entanglement certification for both pure and mixed states. In sum, we clarify the challenges of angle-only certification, while showing that, upon incorporating verifiable external information or supplementary physical constraints, one can still realize practical entanglement criteria independent of $\alpha_\Lambda$ and $\alpha_{\bar{\Lambda}}$, offering methodological guidance for future studies with hyperon pairs and other weakly decaying systems.

\begin{acknowledgments}
J. Pei is supported by the National Natural Science Foundation of China (Project No. 12505121), by the Joint Fund of Henan Province Science and Technology R$\&$D Program (Project No. 245200810077), by the Startup Research Fund of Henan Academy of Sciences (Project No. 20251820001),
and by the Scientific and Technological Research Project of Henan Academy of Sciences (Project No. 20262320001).
L Wu is supported in part by the Natural Science Basic Research Program of Shaanxi, Grant No. 2024JC-YBMS-039.
TL is supported in part by the National Key Research and Development Program of China Grant No. 2020YFC2201504, by the
Projects No. 11875062, No. 11947302, No. 12047503,
and No. 12275333 supported by the National Natural Science Foundation of China, by the Key Research
Program of the Chinese Academy of Sciences, Grant
No. XDPB15, by the Scientific Instrument Developing Project of the Chinese Academy of Sciences, Grant
No. YJKYYQ20190049, by the International Partner
ship Program of Chinese Academy of Sciences for Grand
Challenges, Grant No. 112311KYSB20210012, and by
the Henan Province Outstanding Foreign Scientist Studio Project, No.GZS2025008. 
\end{acknowledgments}

\bibliographystyle{jhep}
\bibliography{jhep}

\end{document}